\documentclass[fleqn,usenatbib]{mnras}

\usepackage{hyperref}
\hypersetup{colorlinks}
\newcommand{\mirisim}{\texttt{MIRISim}}
\newcommand{\miri}{\texttt{MIRI}}
\newcommand{\jwst}{\textit{JWST}}
\newcommand{\fits}{\texttt{FITS}}

\newcommand{\maisie}{\texttt{MAISIE}}

\newcommand{\obssim}{\texttt{OBSSim}}
\newcommand{\skysim}{\texttt{SkySim}}
\newcommand{\SCASim}{\texttt{SCASim}}
\newcommand{\imsim}{\texttt{ImSim}}
\newcommand{\lrssim}{\texttt{LRSSim}}

\usepackage{breakurl}
\newcommand{\python}{\texttt{Python}}
\usepackage{bbding}
\usepackage{graphicx}
\graphicspath{ {figures/} }
\usepackage{xcolor}
\usepackage[normalem]{ulem}
\usepackage{subfig}
\usepackage{dcolumn}
\newcolumntype{d}[1]{D{.}{.}{#1}}
\usepackage{comment}

\usepackage{listings}

\lstdefinestyle{mystyle}{
    commentstyle=\color{codegreen},
    keywordstyle=\color{magenta},
    stringstyle=\color{codepurple},
    basicstyle=\ttfamily\footnotesize,
    breakatwhitespace=false,         
    breaklines=true,                 
    captionpos=b,                    
    keepspaces=true,                 
    numbersep=5pt,                  
    showspaces=false,                
    showstringspaces=false,
    showtabs=false,                  
    tabsize=2
}

\usepackage{natbib}
\title[MIRISim: A Simulator for MIRI on JWST]{MIRISim: A Simulator for the Mid-Infrared Instrument\\ on JWST}

\author[P.~D. Klaassen, V.~C. Geers, et al.]{P.~D. Klaassen$^{1}$,\thanks{Email: pamela.klaassen@stfc.ac.uk} 
V.~C. Geers$^{1}$, 
S.~M. Beard$^{1}$,
A.~D. O'Brien$^{1}$,
C. Cossou$^{2}$,
R. Gastaud$^{3}$,
\newauthor
A. Coulais$^{3,4}$,
J. Schreiber$^{5}$,
P.~J. Kavanagh$^{6}$,
M. Topinka$^{6,7}$,
R. Azzollini$^{8}$,
W. De Meester$^{9}$,
\newauthor
J. Bouwman$^{5}$,
A.~C.~H. Glasse$^{1}$,
A.~M. Glauser$^{10}$,
D.~R. Law$^{11}$,
M. Cracraft$^{11}$,
K. Murray$^{11}$,
\newauthor
B. Sargent$^{11}$,
O.~C.~Jones$^{1}$,
and G.~S. Wright$^{1}$
\\
$^{1}$UK Astronomy Technology Centre, Royal Observatory Edinburgh, Blackford Hill, Edinburgh EH9 3HJ, UK \\ 
$^{2}$Institut d’Astrophysique Spatiale, CNRS/Universit\'e Paris-Sud, Universit\'e Paris-Saclay, b\^atiment 121, Universit\'e Paris-Sud,\\ 91405 Orsay Cedex, France\\
$^{3}$AIM, CEA, CNRS, Université Paris-Saclay, Université Paris Diderot, Sorbonne Paris Cité,F-91191 Gif-sur-Yvette, France\\
$^{4}$LERMA, Observatoire de Paris, CNRS, F-75014, Paris, France\\
$^{5}$Max Planck Institut f\"ur Astronomie, K\"onigstuhl 17,69117 Heidelberg, Germany\\
$^{6}$Dublin Institute for Advanced Studies, School of Cosmic Physics, Astronomy \& Astrophysics Section, 31 Fitzwilliam Place,\\ Dublin 2, Ireland.\\
$^{7}$Department of Theoretical Physics and Astrophysics, Masaryk University, Kotl\'{a}\v{r}sk\'{a} 2, CZ-611 37 Brno, Czech Republic\\
$^{8}$Mullard Space Science Laboratory, University College London,
Holmbury St Mary, Dorking, Surrey RH5 6NT, United Kingdom\\
$^{9}$Institute of Astronomy, KU Leuven, Celestijnenlaan 200D Box 2401, 3001 Leuven, Belgium\\
$^{10}$ETH Zurich, Institute for Particle Physics and Astrophysics, Wolfgang-Pauli-Strasse 27, 8093 Zurich, Switzerland\\
$^{11}$Space Telescope Science Institute, 3700 San Martin Drive, Baltimore, MD 21218, USA}

\date{Accepted 2020 October 28. Received 2020 October 21; in original form 2020 September 21 }
\pubyear{2020}

\begin{document}
\label{firstpage}
\pagerange{\pageref{firstpage}--\pageref{lastpage}}
\maketitle

\begin{abstract}
The Mid-Infrared Instrument (\miri) on the James Webb Space Telescope (\jwst), has imaging, four coronagraphs and both low and medium resolution spectroscopic modes . Being able to simulate \miri{} observations will help commissioning of the instrument, as well as get users familiar with representative data.  We designed the \miri{} instrument simulator (\mirisim{}) to mimic the on-orbit performance of the \miri{} imager and spectrometers using the Calibration Data Products (CDPs) developed by the \miri{} instrument team. The software encorporates accurate representations of the detectors, slicers, distortions, and noise sources along the light path including the telescope's radiative background and cosmic rays. The software also includes a module which enables users to create astronomical scenes to simulate. \mirisim{} is a publicly available \python{} package that can be run at the command line, or from within \python{}. The outputs of \mirisim{} are detector images in the same uncalibrated data format that will be delivered to \miri{} users. These contain the necessary metadata for ingestion by the \jwst{} calibration pipeline.
    
\end{abstract}


\section{Introduction}

The Mid-Infrared Instrument (\miri) on the James Webb Space Telescope (\jwst) observes in the 5-28 $\mu$m wavelength range.

\miri{} is the only Mid-IR instrument on the JWST, and as such, is uniquely capable of probing warm material in the nearby Universe, and to open up new windows at high redshifts for observing the early Universe. A more thorough overview is given in \citet{MIRI1}, however, some of the science cases are summarised below \citep[see also][]{MIRI1}.

\miri{} will probe questions including the origin of exo-planet diversity, as well as how, where, and from what those exo-planets form. With the spectroscopic modes specifically, \miri{} can observe atmospheres of exo-planets, probing their compositions and their potential for hosting life \citep{GTO_1277}. Serendipitously, \miri{} will be able to detect the thermal emission from asteroids, as they cross through its field of view.

To further the  study of protoplanetary (and debris) disks, \miri{} will be used to study the chemistry of the terrestrial planet forming zones of disks, and how the gas in disks evolves as the dust disperses. The high spatial resolution and sensitivity will allow studies of the structure of the thermal emission in these disks, such as depletion zones, rings, rims and even spiral density waves \citep{GTO_1282}.  In the earlier evolutionary stages of these forming stars,  \miri{} will be able to probe the accretion rates onto the stars, and physical structures of the surrounding disks, outflows and envelopes.   \miri{} will probe star formation on all mass scales, from their earliest collapse phase, to how they feed back on to their natal environments \citep{GTO_1290}. This feedback can include the production of photon-dominated regions (PDRs); a warm, primarily neutral region that is being photo-ionised by the UV radiation from the nearby high-mass stars. \miri{}'s high spatial resolution will enable studies of how the material at the edges of PDRs is stratified from the hot ionised material close to the stars to cool embedded material on size scales of $<$ 1$''$ \citep{GTO_1192}.

Beyond the Milky Way, \miri{} will be able to observe and resolve the rings and ejecta of Supernova 1987A, search for the remnant neutron star, and enable the study of the interaction between the ring and the blast wave \citep{GTO_1232}.  In nearby galaxies, \miri{} can be used to identify previously hidden active galactic nuclei, map the stellar kinematics in order to constrain the mass concentration in galaxies, identify and determine the ages of stellar clusters, and understand the spatial structures of the star formation history, variations in the dust extinction and ionisation structures \citep{GTO_1267,GTO_1265}.

At high red-shifts, \miri{} can detect rest frame H$\alpha$ beyond z = 6.7 and Pa-$\alpha$ beyond z=1.6 in galaxies and Quasi-Stellar Objects (QSOs), reaching towards the epoch of re-ionisation. It will observe rest-frame Near-IR emission from evolved stars in dust enshrouded star-forming galaxies at 2$<$z$<$6, to study the obscured star formation and black hole growth in these galaxies \citep{GTO_1263}.

\begin{figure*}
\includegraphics[width=\textwidth]{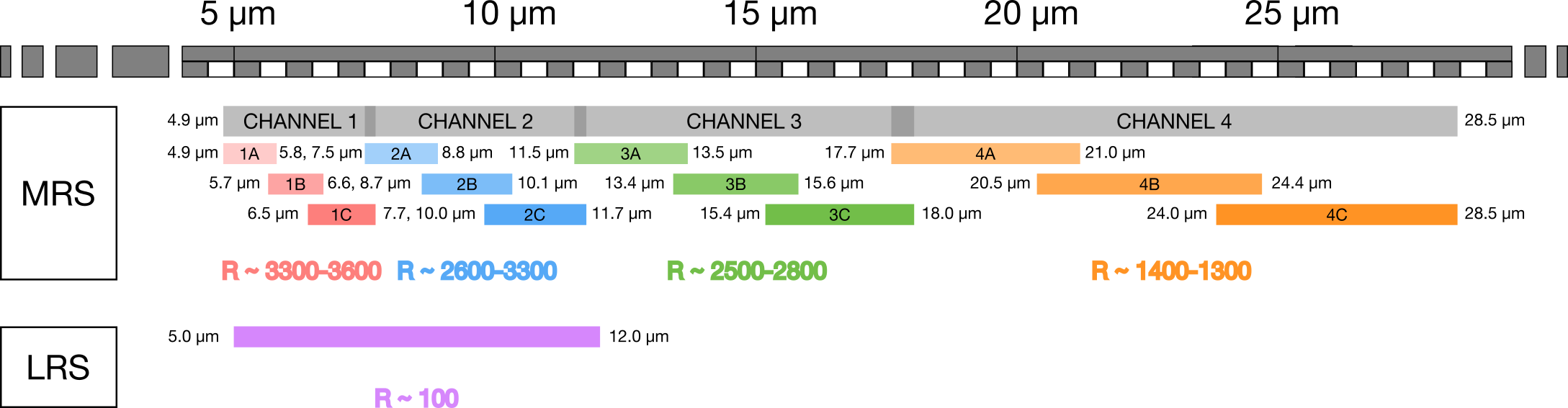}
\caption{Spectral resolution and wavelength coverages for the LRS and each of the \miri{} MRS channels and sub-channels. The full channel ranges of the MRS are listed in the overlapping grey panels towards the top, with the wavelength ranges of each sub-channel are listed below. The purple bar at the bottom shows the wavelength coverage of the LRS.  The listed `R' values denote the spectral resolving power ($\lambda/\Delta\lambda$) for each MRS channel, and the LRS. Image courtesy of N. L\"utzgendorf (ESA). }
\label{fig:spectral_coverage}
\end{figure*}

\begin{figure*}
\includegraphics[width=\textwidth]{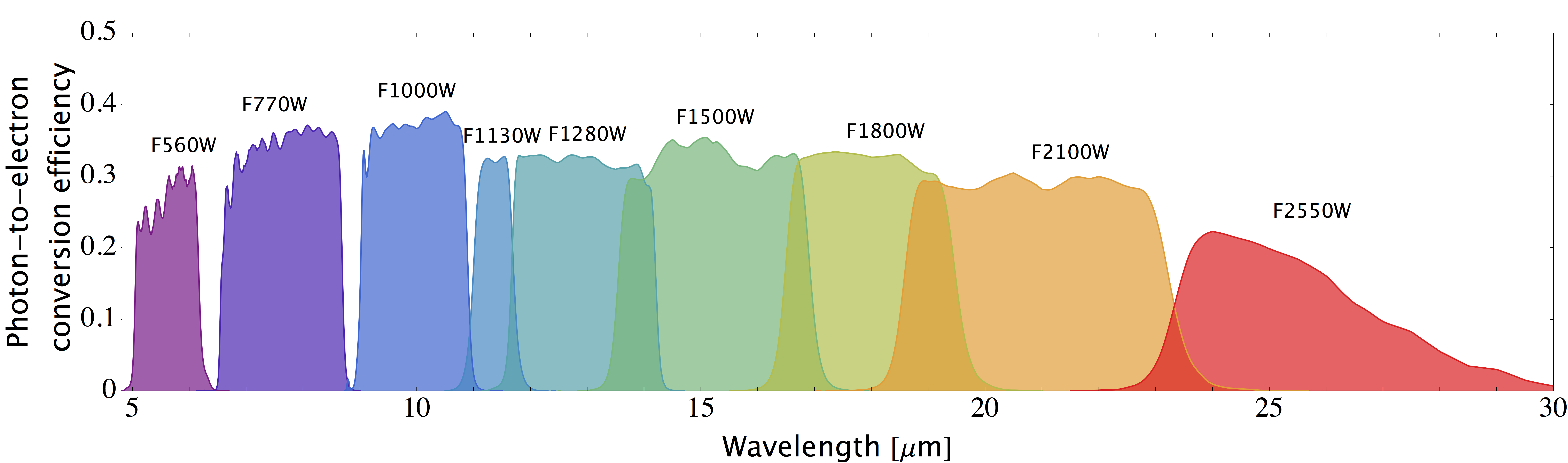}
\caption{Wavelength coverage of the \miri{} imager filters and their photon-to-electron conversion efficiency (PCE) The name of each filter is listed above its range. Their central wavelengths and bandwidths are presented in Table \ref{tab:filters}. Image courtesy of N. L\"utzgendorf (ESA). }
\label{fig:imager_filters}
\end{figure*}

\begin{table}
\begin{center}
\caption{Filter names, central wavelengths and bandwidths of the \miri{} imager filters}
\begin{tabular}{ld{2.1}d{1.1}}
\hline\hline
Filter Name & \multicolumn{1}{c}{$\lambda$} & \multicolumn{1}{c}{$\lambda$/$\Delta\lambda$}\\
\hline
F560W & 5.6 & 5.0\\
F770W & 7.7 & 3.5\\
F1000W & 10.0 & 5.0\\
F1130W & 11.3 & 16.0\\
F1280W & 12.8 & 5.0\\
F1500W & 15.0 & 5.0\\
F1800W & 18.0 & 6.0\\
F2100W & 21.0 & 4.0\\
F2550W & 25.5 & 6.0\\
\hline\hline
\label{tab:filters}
\end{tabular}
\end{center}
\end{table}

By operating in space, above the thermal emission and inherent instability of the atmosphere and by being  cryogenically cooled, \miri{} will deliver unprecedented sensitivities at sub-arcsecond spatial resolutions. The instrument consists of  medium and low resolution spectrographs \citep[respectively]{MIRI6,MIRI4}, an imager \citep{MIRI3}, and four coronagraphs \citep{MIRI5}.  The imager includes 9 filters (see Table \ref{tab:filters}) covering the 5 to 28 $\mu$m spectral range with a field of view of $74''\times113''$ and has the ability to use sub-arrays for bright targets. Within the imager's overall field of view are also three four-quadrant phase mask (4QPM) coronagraphs imaging at 10.65, 11.4 and 15.5 $\mu$m, a Lyot coronagraph at 23 $\mu$m, and the low resolution spectrometer (LRS) covering wavelengths from 5 to 12 $\mu$m with both slit and slitless modes. The medium resolution spectrometer (MRS) is an integral field unit (IFU) spectrometer with its own field of view (ranging from $3.0''\times3.9''$ to $6.7''\times7.7''$ through the full \miri{} wavelength range). 

\begin{figure}
\includegraphics[width=\columnwidth]{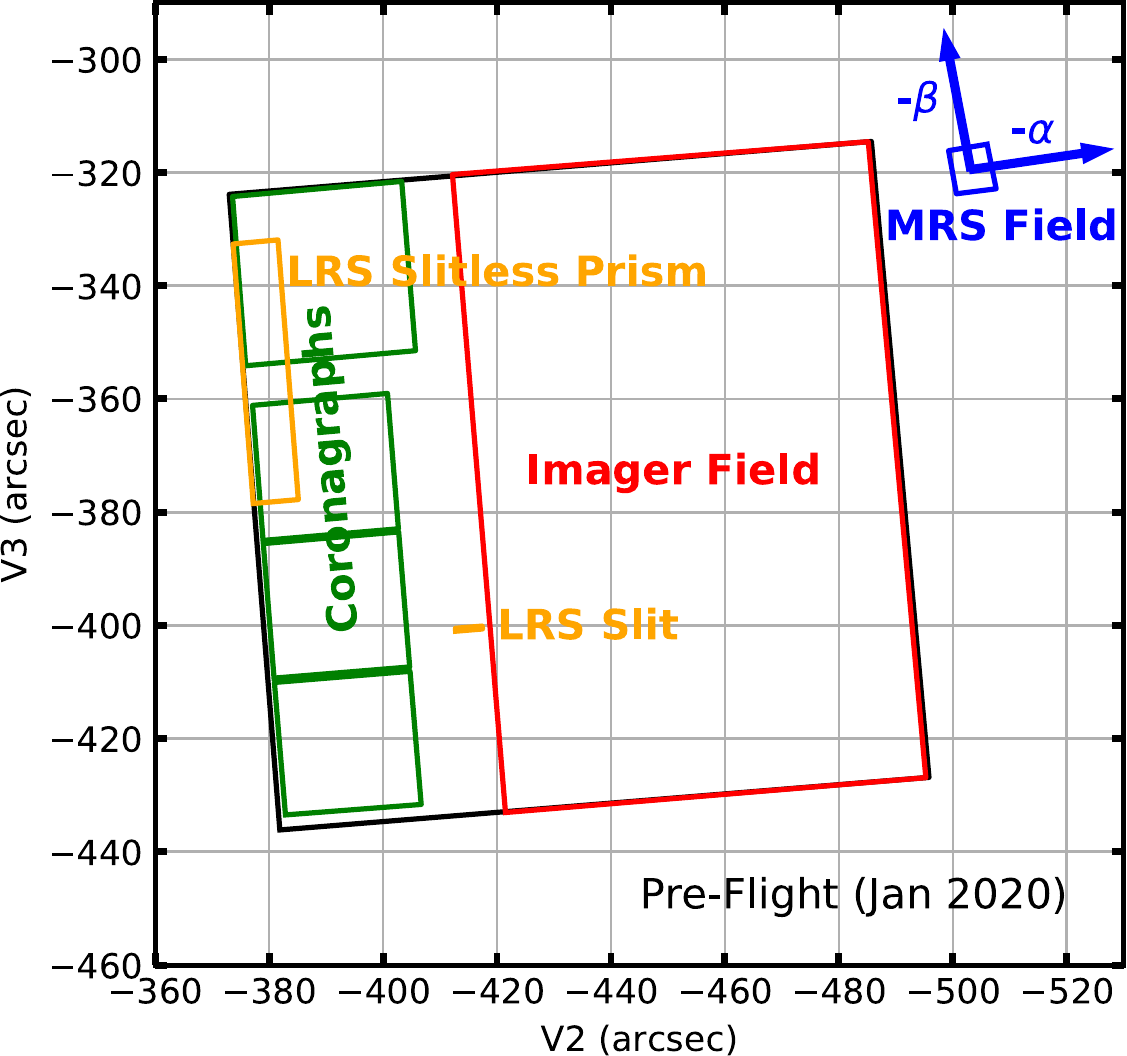}
\caption{Relative positions of the \miri{} components in the \jwst{} focal plane (the v2,v3 coordinate system). The $\alpha$ and $\beta$ axes of the MRS field (in blue) represent the along and across slice directions of the MRS image slicer \citep[see][]{MIRI2}. }
\label{fig:MIRI2_f3}
\end{figure}

\miri{} consists of an optical bench (containing the optics assembly which includes the imager and MRS) and a cooler unit which keeps the system below 7 K. The \miri{} field is selected from the JWST's focal plane by a single pick off mirror, with the regions of the field directed to each optical function as shown in Fig. \ref{fig:MIRI2_f3}, adapted from \citet{MIRI2}. % single pick-off mirror is used for the imager and MRS, with a second fold mirror used for the MRS.  Within the imager, light is then further directed into either the imager, a coronagraph, or the LRS. The relative fields of view of the different \miri{} components are shown in Fig. \ref{fig:MIRI2_f3}, taken from \citet{MIRI2}.  
The spectral resolution and wavelength coverage of both the MRS and LRS are given in  Fig. \ref{fig:spectral_coverage}. For the LRS, the spectral resolution varies from R$\sim$40 to R$\sim160$ at longer wavelengths ($\sim$10 $\mu$m). The MRS wavelength range is covered by four channels, each divided into three sub-bands with the wavelength coverage and specreal resolution shown in Fig. \ref{fig:spectral_coverage}. The coronagraphic and LRS prism filters are  located in the filter wheel and chosen using the same mechanisms as that for the imager filters. The wavelengths of the imaging filters are shown in Fig. \ref{fig:imager_filters}, with additional bandwidth information given in Table \ref{tab:filters}.  The flexibility of the system that combines all of the above allows \miri{} to make significant contributions to a large range of science topics.

\mirisim{} is a \python{} package developed by the \miri{} team to simulate the expected performance of all of \miri's modes except coronagraphy. As such, it is able to create realistic simulations of the kinds of science targets described above. 

Not only can this simulator be used by the commissioning team to verify new calibrations as the on-orbit performance of the instrument is verified, but it can be used by observers wishing to understand whether their proposed observations will allow them to obtain their science goals, or for theorists to show what their predictions would look like if observed with \miri.

\mirisim{} is not alone in being able to be used by the community to better understand how an instrument or telescope works, or to demonstrate the relationship between what can be observed and theorists predictions.  Noteable examples of other simulators which allow the user to simulate their own specific observations include the interferometric simulator for observatories such as ALMA and the JVLA built into \texttt{CASA}\footnote{https://casaguides.nrao.edu/index.php/Guide\_To\_Simulating\_ALMA\_Data}, the ImSim\footnote{https://github.com/LSSTDESC/imSim} and PhoSim\footnote{https://bitbucket.org/phosim/phosim\_release/wiki/Home} packages for the Rubin Observatory, and the MIRAGE\footnote{https://www.stsci.edu/jwst/science-planning/proposal-planning-toolbox/mirage} package for NIRCam and NIRISS instruments.

In this paper, we describe the architecture of \mirisim{} (Section \ref{sec:components}), and how it simulates \miri{} observations. We then give a brief description of how to install and use MIRISim in Section \ref{sec:usage}.  We present the outputs of \mirisim{} in Section \ref{sec:outputs}, and conclude in Section \ref{sec:conclusions}.  %Also included in this paper is an Appendix showing a few examples of how \mirisim{} can be used to simulate some of the MIRI science cases being investigated as part of the Guaranteed Time Observation programmes.

\section{Scope and Limitations}

\mirisim{} was designed to replicate, on a best effort basis, the instrument team's best working knowledge of \miri{}.  It is expected to reproduce the sensitivity of the instrument to within 10-20\% of this `nominal' baseline understanding.  It allows for dithered observations, which are often required for \jwst{} calibration pipeline processing, but will not handle mosaiced observations.  Very little telescope based spatial information is preserved in a \mirisim{} simulation; simulated observations are centered on (RA,DEC) = (0,0), and the simulator does not account for telescope roll angles.  It is based on the Calibration Data Products (CDPs, see below), and as such, instrumental effects that are not properly captured in CDPs are not captured well in the simulator. This includes effects such as the model of spectral fringing through the instrument, since that is still under development by the \miri{} team. It is unclear whether this will be incorporated before launch.

Combined, the above means that science data from the telescope will likely look different from the current understanding available in the simulator.   The results from \mirisim{} should be taken as indicative of \miri{} performance, and users should continue to use the Space Telescope Science Institute (STScI) exposure time calculator (ETC)\footnote{https://jwst.etc.stsci.edu} for a detailed understanding of required exposure times and sensitivities of the observatory.

\subsection{MIRI Calibration Data Products (CDPs)}

% referee based suggestion to re-phrase.
% we need to convey that the CDPs are the products created by the MIRI team to explain the behaviour of the instrument. Since they are the best source for that information, we are using them within MIRISim to simulate the instrument itself.

Where possible, the \miri{} instrument characteristics are parameterised via the use of \miri{} CDPs, which are a set of deliverables to STScI. These data products are the best representation of the instrument, and form the basis of the \miri{} calibration reference data system (CRDS) files used in the \jwst{} calibration pipeline\footnote{https://jwst-pipeline.readthedocs.io/en/latest}.

In addition to delivering them to STScI, the \miri{} team uses these data products  within \mirisim{} to characterise the instrument. As CDPs are updated, and new versions incorporated into the pipeline, they are also incorporated into \mirisim{} to provide the most accurate representation of the instrument the team can provide. Whenever there is a change in an underlying CDP, a note is made in the \mirisim{} release notes explaining the versions updates.

\section{\mirisim{} components}
\label{sec:components}

\begin{figure*}
\includegraphics[width=\textwidth]{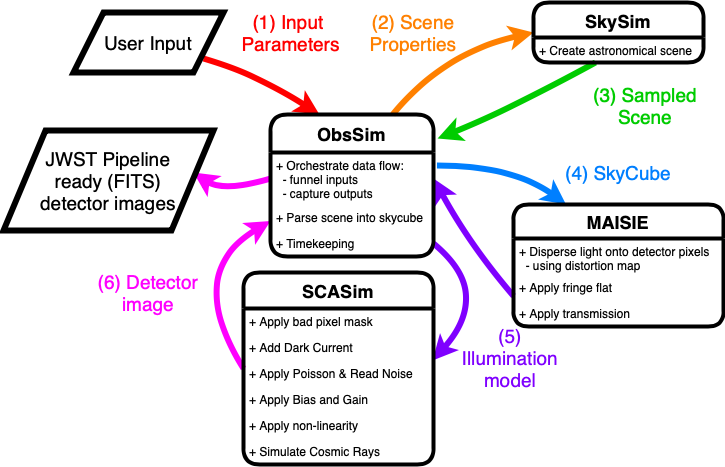}
\caption{Overview of the path data takes through the MRS version of \mirisim{}.  The user supplies input parameters (either at the command line, or from within \python{}), and \obssim{} orchestrates the creation of the scene, the dispersion through \maisie{}, sending the output illumination model to \SCASim{}, and creating a FITS file of the resultant detector image.}
\label{fig:MRS_overview}
\end{figure*}

\begin{figure*}
\includegraphics[width=\textwidth]{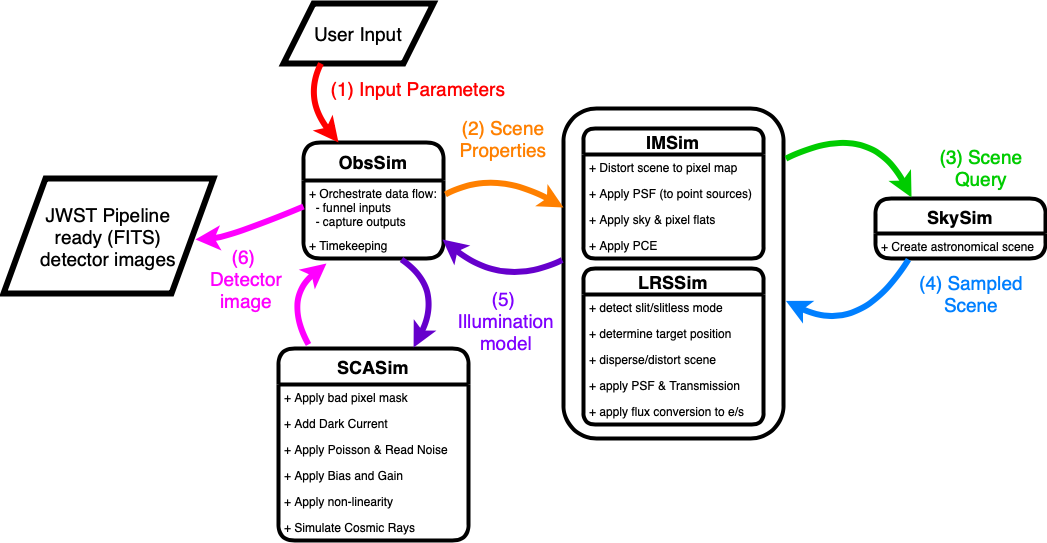}
\caption{Overview of the path data takes through the imager or LRS versions of \mirisim{}.  The user supplies input parameters (either at the command line, or from within \python{}), and \obssim{} sends the scene parameters to \lrssim{} or \imsim{} to create a scene and determine the illumination model. \obssim{} then takes that illumination model (as it does in the MRS case) and passes it to \SCASim{} to create the final detector image, which \obssim{} then turns into  a FITS file ready for ingest into the \jwst{} calibration pipeline.}
\label{fig:IMA_overview}
\end{figure*}

%Gillian's suggested addition:
%Suggest adding: A modular approach to the design of MIRISim was chosen in order to build flexibility to incorporate increasing detail of the instrument characteristics as hardware was tested.  The design of the simulator reflects the modular hardware approach and commonality of the detector characteristics for the imager and spectrometer modules. 

\mirisim{} is a \python{} package consisting of a number of distinct components, some of which simulate the various light paths through the instrument (\imsim{}, \miri{}-\maisie{} and \lrssim{}), while others orchestrate (\obssim{}), create astronomical scenes (\skysim{}), and process detector effects (\SCASim{}) for the simulation.  A modular approach to the design of \mirisim{} was chosen in order to build flexibility to incorporate increasing detail of the instrument characteristics as hardware was tested.  The design of the simulator reflects the  modular approach to the hardware and the use of common detectors between the imager and spectrometers.  It also facilitates re-use of software modules to build simulators for future instruments.   The user interface to \mirisim{} is described in Section \ref{sec:usage}, but here we describe the individual components that make up the simulator. 

For reference, the path through the \mirisim{} components for MRS and imager/LRS simulations are given (respectively) in Figures \ref{fig:MRS_overview} and \ref{fig:IMA_overview}. In these figures, the flow of data through \mirisim{} starts with users inputting a reference astronomical scene to be simulated, and the observation parameters for the simulation which follow the nomenclature used in the \jwst{} Astronomer's Proposal Tool (APT)\footnote{https://www.stsci.edu/scientific-community/software/astronomers-proposal-tool-apt}. The orchestration module (\obssim{}) then parses the scene and simulation parameters and sends the commands to the relevant modules which create the sampled scene (or sky cube) and then create a model of how the detector would be illuminated (creating an illumination model) this output  and the simulation parameters are then passed to the Sensor Chip Assembly simulator (\SCASim{}) where the exposures are calculated, and the final detector images are created.  Thus, the final product of a \mirisim{} simulation, regardless of light path, is a \texttt{detector image}.   For a description of the detector images, see Section \ref{sec:SCASim}. ObsSim then wraps the detector images into FITS files with the correct header information so that files can be read by the \jwst{} calibration pipeline.

Below, we describe the individual \mirisim{} components in greater detail.

\subsection{\skysim{}}

The SkySim module creates the astronomical `scene' used by \mirisim{} to simulate the observation with the use requested instrument setup. A `scene' here is defined as a collections of objects placed within the instrument field of view with spatial and spectral properties which can be mapped into illumination models for the detectors.  We use the same terminology here as in the \jwst{} Exposure Time Calculator (ETC), but allow for more flexibility in the definition of objects within a scene.  The user can build up a set of objects to represent what the sky is expected to look like with \miri{}.

These `scenes' can take the form of a series of point and extended `sources' (emitters in the field of view) specified by the user with positions, extents and shapes (for extended objects), and spectral energy distributions (SEDs) of various types. There is also the option to give ascii files containing SEDs for the individual components of the scene.  \skysim{} adds a `background' component  to user defined scenes, which consists of the non-negligible telescope thermal background \citep[see][]{MIRI9} as well as other effects such as zodiacal light.  It can also read \texttt{pysynphot} SED libraries, which are downloaded as part of the installation process (see Section \ref{sec:installation}) which can be used to populate point or extended objects with library SEDs providing that the SEDs are valid across the required \miri{} wavelength range.

\skysim{} can also interpret user-provided \fits{} cubes as ready-made input scenes, only requiring a minimal set of \fits{} header keywords to parse them. Most standard \fits{} keywords are accepted, with the exception that the wavelength axis must be specified in microns, not Angstroms. Warnings are given if the positional coordinates are not specified properly.

\subsection{\obssim{}}

\obssim{} is the \mirisim{} module that orchestrates the rest of the components. It reads in the inputs (via a configuration parser), and ultimately creates the \fits{} file outputs of each stage of the simulation. It sets up dither positions, coordinates the number of exposures, and keeps track of the timing of the simulated observation to ensure that the header metadata is realistically tracking the time it takes for an observation.

%\pk{ObsSim nominally tracks the time so that the exposure start time in the header information in each dither location is realistic}

% in the constants module: wait time for fine steering mirror, and wait time for Small angle manoeuvre
% asking for multiple exposures, we add an extra wait second for setting up the next exposure

%\pk{SCAsim doesn't need the obssim clock to calculate the exposure duration to determine the cosmic ray stats. it does that internally}

%\pk{the goal of the communication of time from ObsSim to SCASim was for latents. but since we don't deal with latents, this isn't something that was implemented since latents decay on short enough timescales that we don't worry about it.}

For the MRS light path, \obssim{} has the added task of creating a skycube by griding the components of the scene created in \skysim{} onto a cube object which can then be read by \miri{}-\maisie{}.

\obssim{} is also responsible for processing the illumination model outputs of the three light path simulator components (ImSim, LRSSim or \miri{}-\maisie{}) and passing them to the sensor chip assembly simulator (SCASim) along with the observing parameters (number of groups, integration and exposures, whether to simulate SLOW or FAST mode, etc) input by the user to create the detector image which is the end product of \mirisim{}. Each of these components are described in more detail below (See Sections \ref{sec:maisie} - \ref{sec:SCASim}).

\subsection{\miri{}-\maisie{}}
\label{sec:maisie}

\miri{}-\maisie{} is built using the  Multipurpose Astronomical Instrument Simulator Environment \citep[\maisie{};][]{MAISIE} which has been tailored to the \miri{}-MRS light path.  Its components are based on those of SpecSim \citep{Specsim}, and incorporates the transmission, distortion, dispersion and both the results of the point and line spread function CDPs for MRS observations computed in \obssim{}.  The output of \miri{}-\maisie{} is an illumination model (from which \obssim{} makes an output \fits{} file) for each exposure covering the specified (i.e.  A, B, and C as shown in Fig. \ref{fig:spectral_coverage}) sub-channel of either the `Short' (Ch1-2, 4.8 to 11.9 $\mu$m) or `Long' (Ch3-4, 11.5-28.8 $\mu$m) wavelength arms of the MRS, or of both the `Short' and `Long' arms simultaneously.  The illumination models are then used by \SCASim{} to create detector images. The bottom right panel of Fig. \ref{fig:MRSoutputs} shows such a detector image for Channels 1a and 2a.

\subsection{\imsim{}}
\label{sec:imsim}

\imsim{} simulates the light path through the primary imager field of view (i.e. not the LRS or coronagraphs).  It queries \skysim{} to create the astronomical scene to be simulated, determines the FOV to be imaged (i.e. whether to image the full frame or a sub-array), applies the requested filter, distortion sky-flat and photon conversion efficiency (PCE) CDPs, and convolves the scene components with the point spread function (PSF). The resultant illumination model is an intermediate product of \mirisim{} (written out to a \fits{} file by \obssim{}), which is fed (via \obssim{}) to \SCASim{} to create a detector image.  As seen in Figure \ref{fig:IMoutputs}, \mirisim{} illumination models and subsequent detector images do show light passing through the coronagraphic fields of view, despite not doing any coronographic simulation.

During instrument ground testing, a cruciform pattern was seen in the PSF from internal scatterings in the detector. This pattern has been incorporated into the PSF CDPs, and so is not accounted for separately in \mirisim{}. The cruciform is only applied when the target is in field because it is a detector effect.

\subsection{\lrssim{}}
\label{sec:lrssim}

The LRS has two modes: slitted and slit-less. \lrssim{} simulates both of these modes, dispersing the light from the required part of the simulated `scene'. Slitless mode is to be used for point sources, and when used in this mode, only the SLITLESSPRISM sub-array (located towards the top left of the detector. In slit mode, the full imager field of view is simulated. The positions of the targets on the detector are shown in Figure 2 of \citet{MIRI4}.  Similarly to \imsim{}, \lrssim{} applies transmission, distortion, PCE (or optional spectral response) and PSF CDP (or, optionally the webbPSF model) to the dispersed data to produce an illumination model, which \obssim{} then delivers to \SCASim{} to create the final detector images. To support commissioning but also normal observations, LRSSim can optionally simulate the complete detector array to identify possibly disturbing (extended) nearby sources outside the slit or slitless area.

\subsection{\SCASim{}}
\label{sec:SCASim}

The electronic (pixel gain, dark current and pixel response) and detector (reset anomaly, last frame effect, droop and drift) effects  of the Sensor Chip Assembly (SCA) itself are described in detail in \citet{MIRI8}.  The SCAs for the imager/LRS and MRS detectors are identical \citep[as described in greater detail in ]{MIRI8}, and as such, the outputs from \maisie{}, \lrssim{} and \imsim{} can all be processed by the same simulator module.

The Sensor Chip Assembly simulator \citep[\SCASim{};][]{SCASim} is the final component  of \mirisim{}. It simulates the most important \miri{} detector effects of the instrument. These include known cosmetic effects (e.g. bad pixels), detector dark current,  electronic non-linearity effects, the pixel flat-field,  noise sources (e.g. read noise, shot noise) and a plug-in cosmic ray event library (based on Robberto, M., 2010, priv. comm) which can be updated as our knowledge of these effects improves. The current treatment of cosmic rays is based on the interpixel capacitance corrections of the \miri{} detectors as defined by Gaspar et al (2007, priv. comm), with refinements subsequently suggested by G. Rieke (priv. comm).

\SCASim{} samples the illumination model based on the number of groups, integrations and exposures requested, applying the noise sources as appropriate.  It then returns 3D \texttt{detector images} with  slices corresponding to the individual groups for each integration (and sets of images when there are multiple exposures). The detector images represent the sensitivity of the instrument to light, and reproduce, to the teams best understanding, the defects, non-linearities and noise within the array.  These files, when converted to FITS format by \obssim{}, and given the appropriate header information, can be read into the \jwst{} calibration pipeline as Level 1B datasets.

\section{Getting and Using \mirisim{}}
\label{sec:usage}

\subsection{Installation}
\label{sec:installation}
\mirisim{} has been developed and tested on Apple (macOS) and Linux systems, and  the minimum system requirements for running \mirisim{} are comparable to those for the \jwst{} calibration pipeline.  The code itself is \python{} 3 based, and \mirisim{} has been bundled into an Anaconda\footnote{https://www.anaconda.com} environment to ease installation.

Detailed installation instructions for \mirisim{} can be found on the MIRICLE  website\footnote{http://miricle.org}. From this website, users can download an installation script.  This self-updating bash script will always download the most recent version of \mirisim{} and install it in a new anaconda environment.   The primary environment variables to be set are the location of the Calibration Data Products (CDPs), the PySynPhot libraries, and where \mirisim{} is being installed.

\subsection{Usage}

There are two ways of interacting with \mirisim{}: at the command line, and from within \python{} itself.  Either option can accept (and for command line, requires) user formatted input data files to specify how the simulation should be setup, along with a description of the astronomical scene to simulate.  These options are described in more detail below.

\subsection{User Input Files}

Whether working at the command line, or from within \python{}, \mirisim{} takes a number of input files which define parameters that define how the astronomical scene is to be setup, what kind of simulation to run, and a number of advanced features (primarily for use in \jwst{} commissioning).   These files, and their formatting are described in more detail in the \mirisim{} Users Guide\footnote{\burl{http://miri.ster.kuleuven.be/bin/view/Public/MIRISim_Public}}. The input files can be used at both the command line, or as inputs into \python{} runs of the simulator. Through the rest of this text, we refer to the scene, simulation and simulator files with these template names. The scene file (e.g. \texttt{scene.ini}) houses the specification of the astronomical scene to be simulated (e.g. a point source 0.5$''$ from the field center with a blackbody SED), and \texttt{simulation.ini} houses the properties of the simulation itself (e.g. which imager filter to use, how many exposures to do, how many groups should be in each integration, etc).

When running \mirisim{} at the command line, at least one input file is required  (\texttt{simulation.ini}), with the understanding that the \texttt{scene.ini} file is specified within the \texttt{simulation.ini}, that it exists, and can be used. Alternatively, a scene file can be specified at run time by simply adding the name of the scene file to the end of the above command. This would override any file named in the simulation input file.

Within \python{} a scene and a simulation can be built up in a much more dynamic way, by importing the various \mirisim{} components, as described in more detail in the User Guide. 

\subsection{Formatted input files}

The formatting associated with the input files for \mirisim{} takes the same form as those to be used for the JWST Calibration pipeline. An example of the formatting is shown below for setting up the MRS portion of the simulation file:

\lstset{language=csh}
%\lstset{frame=single}
\lstset{basicstyle=\ttfamily\footnotesize}
\begin{lstlisting}[]
[Integration_and_patterns]

  [[MRS_configuration]]
    disperser  = SHORT                  
    detector   = SW                    
    Mode       = FAST                 
    Exposures  = 2                    
    Integrations = 2                  
    Frames     = 100                  
\end{lstlisting}

\noindent The input files consist of sets of named nested groups, with individual properties (e.g. number of integrations to simulate) set within the nested structure provided by the number of square brackets in the section headings.

\section{Results}
\label{sec:outputs}

The results of a \mirisim{} simulation are output as FITS files in the same uncalibrated data format which will be delivered to \miri{} users, i.e. \mirisim{} results are in the same format as real \jwst{} data, and are readable by the \jwst{} calibration pipeline as 3D ramp data (\jwst{} uncalibrated data format). In accordance with this, the science images are stored in the \texttt{SCI} header extension by default.

\mirisim{} creates an output directory in the current working directory with a pathname that contains the date and time the simulation was started (e.g. \texttt{yyyymmdd\_hhmmss\_mirisim}). In that directory,  copies are made of the \texttt{scene.ini} and \texttt{simulation.ini} files used to drive the simulation (or created if the simulation was manually built within python), and a number of subdirectories are created depending on the type of simulation performed. For all optical paths, directories are created for intermediate product illumination models (\texttt{illum\_models}) and final product detector images (\texttt{det\_images}).  The illumination models are the outputs of \imsim{}, \lrssim{} or \miri{}-\maisie{}, which show how the detector is to be illuminated. The detector images are the final outputs of \mirisim{}: the illumination models that have been processed through \SCASim{}, and saved to \jwst{} calibration pipeline ready \fits{} files.  Examples of the illumination models and detector images for the imager, MRS, and LRS are given in Figures \ref{fig:IMoutputs}, \ref{fig:MRSoutputs} and \ref{fig:LRSoutputs} (respectively).

For MRS simulations, a third directory is created for datacubes (\texttt{skycubes}). These cubes are used by \miri{}-\maisie{} to disperse light into the illumination models. The  skycubes can also help the user understand what their astronomical scene looked like before the light was dispersed.

\subsection{Example \mirisim{} simulations }

\subsubsection{Imager Example}
\begin{figure*}
\centering
\includegraphics[width=\textwidth]{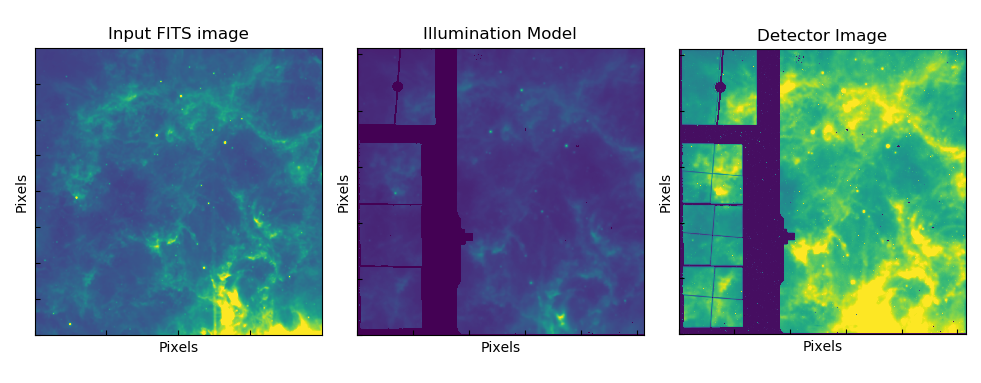}
\caption{Imager outputs from a 100 frame simulation using filter F1130W (11.3 $\mu$m filter), and a WISE image of Carina interpolated to the right wavelength, and moved to a distance of 50 kpc.}
\label{fig:IMoutputs}
\end{figure*}

Fig. \ref{fig:IMoutputs} shows an example of an imager simulation.  The input for the simulation was an interpolation of data from bands 2 and 3 of the \textit{WISE} data of the Carina nebula \citep{2010AJ....140.1868W}.  A linear interpolation of the 4.6 and 12.2 $\mu$m data was created to simulate 11.3 $\mu$m observations, and the spatial resolution information in the header was modified to represent higher resolution data (i.e the \texttt{cdelt} parameters were changed from $\sim$1.37$''$ to $\sim$0.06$''$).  The resultant FITS file is shown in the left panel of Fig. \ref{fig:IMoutputs}. The middle panel shows the model of the illumination presented to the detector through the imager field of view.  The right panel shows the resultant final data product of an imager simulation. The simulation parameters used here consisted of using F1130W with the full imager field of view to observe 100 groups in a single integration using FAST mode. The detector image shown here is the second last group/frame captured in the simulated 3D ramp data at a single dither position.

\subsubsection{MRS Example}

\begin{figure*}
\centering
\includegraphics[width=0.9\textwidth]{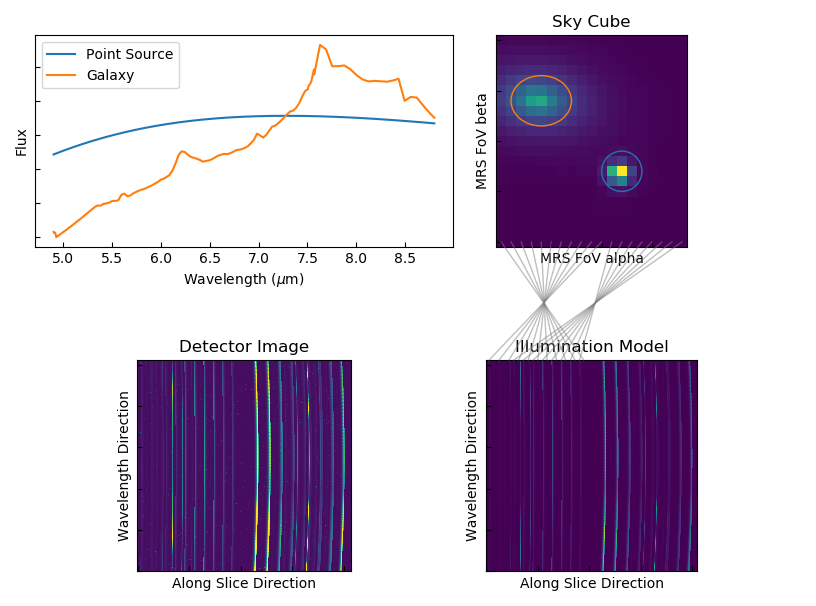}
\caption{MRS outputs from a 100 frame simulation showing {\bf Top Left:} Spectra of the two targets (a point source and an extended ellipse with  blackbody and galaxy spectra, respectively). {\bf Top Right:} Locations of the two targets in the MRS field of view. {\bf Bottom Right:} Model of how the detector is being illuminated by the scene. The grey lines between the top and bottom right panels indicate how the slices in Channel 1A are mapped to the left portion of the detector. {\bf Bottom Left:} Final detector image (in JWST Level 1B data format) of the simulation. }
\label{fig:MRSoutputs}
\end{figure*}

Fig. \ref{fig:MRSoutputs} shows an example of an MRS simulation.  Here, the input is built up of individual components created within \skysim{}.  The first is an extended object (elliptical galaxy) with an effective radius of 1$''$, a Sersic index of 1, and an axial ratio of 0.5 placed at a position angle of 85 deg. The spectral energy distribution (SED) of NGC 1068 was then applied to the extended object. The SED was populated using the Brown  atlas within \texttt{pySynPhot} \citep{Brown14}. The Galaxy and its SED are highlighted in orange in the top two panels of Fig. \ref{fig:MRSoutputs} which show the spectral (left) and spatial (right) distributions of the objects.  The other target placed in the MRS field of view was a point source with a blackbody SED and characteristic temperature of 700K. This target, and its SED are highlighted in blue in the top two panels of Fig. \ref{fig:MRSoutputs}.  As with the imager above, the example MRS simulation consisted of 100 integrations in a single exposure, however here we used SLOW mode readouts. Only the `SHORT' wavelength arm (Channels 1 and 2) was simulated, and all results here are for the `A' sub-channel. This means that the images in the bottom panels of Fig. \ref{fig:MRSoutputs} show the data from 1A and 2A only, with 1A shown in the left half of each figure, and 2A shown in the right.  The grey lines between the skycube (top right) and illumination model (bottom right) panels show how the image slices are mapped to the detector in Channel 1.  Similar lines could be drawn for ch2a (to map to the right portion of the illumination model), but were left out for clarity.

\subsubsection{LRS Slitless Mode Example}

\begin{figure}
\centering
\includegraphics[width=\columnwidth]{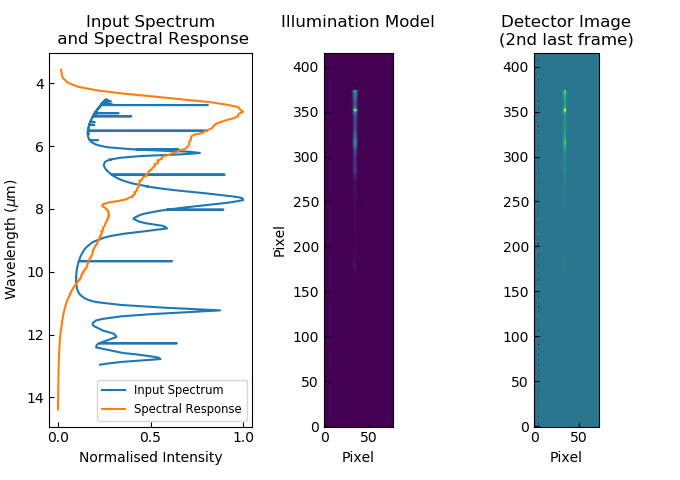}
\caption{LRS slitless outputs for a 100 frame simulation showing a point source with a spectrum of a photon dominated region (PDR).}
\label{fig:LRSoutputs}
\end{figure}

Fig. \ref{fig:LRSoutputs} shows the outputs of an LRS slitless simulation in which a point source was placed at the `source' position for the slitless spectrograph \citep[See Figure 2 of ][]{MIRI4}.  It was given the photon dominated region spectrum shown in blue in the left hand panel of Fig. \ref{fig:LRSoutputs}, and the resultant illumination model is shown in the middle panel.  The LRS spectral response is also shown (in orange) in the left panel of the figure to show why the longer wavelength spectral features of the PDR spectrum are not as prominent in the spectral trace provided in the middle panel.  The right panel of Fig. \ref{fig:LRSoutputs} shows the 2nd last group/frame of a the 3D detector image resulting from a single integration of 100 groups using the FAST readout mode. The imager filter wheel was set to the P750L filter, and the sub-array to readout was set to \texttt{SLITLESSPRISM}. 

\section{Conclusions}
\label{sec:conclusions}

\mirisim{} is the simulator developed by the \miri{} instrument team which reproduces the expected on-orbit performance of the instrument. It simulates each of the light paths through \miri{}  with the exception of coronagraphy. It is built in \python{}, and comes bundled in an Anaconda environment which incorporates all of its dependencies.  It can be used both at the command line, or interactively in a \python{} shell.  It draws on the calibration data products (CDPs) developed by the \miri{} team as we characterise the instrument.

The products of \mirisim{} are \fits{} files with valid metadata and header information that allow them to be ingested into the \jwst{} pipeline as if they were true \miri{} data.

\mirisim{} is expected to be used for various purposes by a broad range of communities. For instance, it will be used by the instrument team during commissioning to correlate observations with expected performances and help refining our understanding of the instrument. It can be used by prospective \jwst{} users to understand the noise characteristics they should expect when imaging faint targets, while others may find it useful to familiarise themselves with data formats from the instrument, to aid in developing observing plans, or for producing realistic inputs when training in data processing and analysis using the \jwst{} calibration pipeline. It can also be used by modellers who want to show how their model predictions could be tested with \miri{} observations.

\section*{Acknowledgements}
The authors would like to thank the anonymous referee for their suggestions which have improved the clarity of this manuscript. The acknowledgements were compiled using the Astronomy Acknowledgement Generator. This research made use of Astropy, a community-developed core Python package for Astronomy \citep{2018AJ....156..123A, 2013A&A...558A..33A} This publication makes use of data products from the Wide-field Infrared Survey Explorer\citep{2010AJ....140.1868W}, which is a joint project of the University of California, Los Angeles, and the Jet Propulsion Laboratory/California Institute of Technology, funded by the National Aeronautics and Space Administration. 

\section*{Data Availability}
No new data were generated or analysed in support of this research.  Instructions for downloading the \mirisim{} code can be found in Section \ref{sec:usage}.

\bibliographystyle{mnras}
\bibliography{main_resubmit}

% Don't change these lines
\bsp	% typesetting comment
\label{lastpage}
\end{document}